\documentclass[conference,a4paper]{IEEEtran}
\usepackage{amsmath,amsfonts}
\usepackage{algorithmic}
\usepackage{algorithm}
\usepackage{array}
\usepackage{textcomp}
\usepackage{stfloats}
\usepackage{lettrine}

\usepackage{url}
\usepackage{subcaption}
\usepackage{verbatim}
\usepackage{graphicx}
\usepackage{cite,flushend}
\usepackage{algorithm,algorithmic,amsmath,amssymb,amsthm,bbm,cite,color,graphicx,microtype,url}
\usepackage[USenglish]{babel}
\usepackage{hyperref}
\usepackage[utf8]{inputenc}
\usepackage[T1]{fontenc}

\newcommand{\herm}{\mathrm{H}}

\newtheorem{remark}{Remark}

\IEEEoverridecommandlockouts
\begin{document}

\title{On the Detection of Non-Cooperative RISs: Scan $B$-Testing via Deep Support Vector Data Description}

\author{
   \IEEEauthorblockN{George Stamatelis$^1$, Panagiotis Gavriilidis$^1$, Aymen Fakhreddine$^2$, and George C. Alexandropoulos$^{1,3}$}
\IEEEauthorblockA{$^1$Department of Informatics and Telecommunications, National and Kapodistrian University of Athens, Greece}
\IEEEauthorblockA{$^2$Institute of Networked and Embedded Systems,
University of Klagenfurt, Austria}
\IEEEauthorblockA{$^3$Department of Electrical and Computer Engineering, University of Illinois Chicago, USA
\\
e-mails: \{georgestamat, pangavr, alexandg\}@di.uoa.gr, aymen.fakhreddine@aau.at
}
\thanks{This work has been supported by the SNS JU projects TERRAMETA and 6G-DISAC under the EU's Horizon Europe research and innovation programme under Grant Agreement numbers 101097101 and 101139130, respectively. 
}
}

\maketitle
\begin{abstract}
In this paper, we study the problem of promptly detecting the presence of non-cooperative activity from one or more Reconfigurable Intelligent Surfaces (RISs) with unknown characteristics lying in the vicinity of a Multiple-Input Multiple-Output (MIMO) communication system using Orthogonal
Frequency-Division Multiplexing (OFDM) transmissions. We first present a novel wideband channel model incorporating RISs as well as non-reconfigurable stationary surfaces, which captures both the effect of the RIS actuation time on the channel in the frequency domain as well as the difference
between changing phase configurations during or among
transmissions. Considering that RISs may operate under the coordination of a third-party system, and thus, may negatively impact the communication of the intended MIMO OFDM system, we present a novel RIS activity detection framework that is unaware of the distribution of the phase configuration of any of the non-cooperative RISs. In particular, capitalizing on the knowledge of the data distribution at the multi-antenna receiver, we design a novel online change point detection statistic that combines a deep support vector data description model with the scan $B$-test. The presented numerical investigations demonstrate the improved detection accuracy as well as decreased computational complexity of the proposed RIS detection approach over existing change point detection schemes.
\end{abstract}


\section{Introduction}
Reconfigurable Intelligent Surfaces (RISs), comprising numerous metamaterials with dynamically tunable electromagnetic responses~\cite{RISsurvey2023}, constitute ultra-lightweight planar structures that can be used to coat building facades, room walls, or vehicles, and are recently considered as one of the candidate technologies for the next generation of wireless networks~\cite{6Gdisac}. They can enable over-the-air signal propagation programmability in an energy-efficient manner, thus, transforming wireless channels to software-defined entities~\cite{AlexandroStylianoPervasive} that can be optimized for various objectives, e.g., enhanced multi-user connectivity~\cite{over_the_air_RIS}, localization~\cite{RIS_loc}, and integrated sensing and communications~\cite{RIS_isac}. 

The core features of RISs, with more pronounced their low hardware footprint recently including even almost transparent designs, have been also lately leveraged for eavesdropping and jamming~\cite{RIS_safeguarding,RIS_Threat_OJCOM,WirelessCommIllegalRIS,metasurf_manip_onleg,RIS_Reciprocity,USE_IRS}. RIS-enabled threat models and proactive countermeasure designs of active and reflective beamforming as well as artificial noise were presented in~\cite{RIS_safeguarding} and~\cite{RIS_Threat_OJCOM} for different knowledge levels of the eavesdropper's channel and unawareness of the presence of a malicious RIS. Jamming attacks based on RISs that are difficult to detect were presented in \cite{RIS_Reciprocity}. Scenarios where the adversary takes control of a legitimate RIS were discussed in~\cite{metasurf_manip_onleg}, while \cite{WirelessCommIllegalRIS} studied the case where an adversary places an RIS in the vicinity of a communication pair for information leakage~\cite{WirelessCommIllegalRIS}. The potential of RISs to decrease the received signal strength in a legitimate link, while enhancing it towards an eavesdropper, was analyzed in~\cite{USE_IRS}. However, none of the latter works focused on detecting the presence of malicious RISs in environments where legitimate communications take place, which could then trigger efficient reactive legitimate physical-layer designs.

On the other hand, as described in one of the RIS deployment use cases presented in~\cite{RIS_challenges}, metasurfaces can be deployed from a single cellular operator in an area of intended coverage, where one or many other operators are also active. Those RISs should have a carefully designed bandwidth of influence to provide reconfigurable reflections to the owner operator, while leaving unaltered the signals spanning the bandwidth allocated to the other operator(s). The latter implies that those RISs should react similar to the surface material that hosts them, upon impinging signals from any of the unintended operators. However, such an RIS design for very closely allocated frequency bands is hard to achieve with up-to-date hardware technologies~\cite{RIS_THz_terrameta}. To this end, an approach that detects third-party non-cooperating RISs would help operators to sense relevant unwanted activity provoking countermeasure designs.  

Motivated by the latter two applications of unwanted RIS operations, we focus, in this paper, on the problem of detecting the operation of one or more non-cooperative RISs in the vicinity of point-to-point Multiple-Input Multiple-Output (MIMO) communication systems. Capitalizing on the discrete-time multipath channel model of~\cite{delay_d_channel} and the coupled-dipole formalism of RIS-parametrized channels of~\cite{PhysFad}, we first introduce a wideband channel model incorporating RISs as well as non-reconfigurable stationary surfaces. 
Then, we present a novel formulation of the RIS activity detection problem as an online sequential change point detection problem~\cite{changePointSurveys}. Considering that the characteristics of any non-cooperative RISs in the wireless environment of interest are unknown to the intended MIMO OFDM communication system, we design a novel online, distribution-free change point detection statistic that combines a deep Support Vector Data Description (dSVDD)~\cite{svdd,dsvdd} model with the scan $B$-test~\cite{scanB}. Our extensive simulation results for the considered RIS detection problem showcase the superiority of our detection approach over existing change point detection methods, both in terms of detection accuracy as well as computational complexity.


\textit{Notations:}
Lower case bold letters refer to vectors, e.g. $\mathbf{x}$, and upper case bold letters indicate matrices, e.g. $\mathbf{X}$. Calligraphic letters, e.g., $\mathcal{X}$, are reserved for sets and $E[\cdot]$ denotes the expectation operator and ${\rm Pr}[\cdot]$ returns the probability. $\mathbf{0}_{N \times M}$ and $\mathbf{I}_{N }$ denote the \(N \times M\) zero matrix and \(N \times N\) identity matrix, respectively. 
$\mathbb{R}$ and $\mathbb{C}$ are the sets of the real and complex numbers. Finally, $\jmath\triangleq\sqrt{-1}$ is the imaginary unit and $x\sim\mathcal{CN}(0,\sigma^2)$ represents a complex normal random variable with zero mean and variance $\sigma^2$.

\section{Channel Model and Problem Formulation}
We consider a mobile User Equipment (UE) possessing an \(N_{\rm ue}\)-element Uniform Linear Array (ULA) that wishes to communicate in the uplink direction with a Base Station (BS), which is equipped with a Uniform Rectangular Array (URA) comprising \(N_{\rm bs}\) antenna elements. Orthogonal Frequency Division Multiplexing (OFDM) transmission is adopted with \(K\) subcarriers and a cyclic prefix of length \(D\) per OFDM symbol. In the wireless propagation environment of interest, there might exist RISs that have not been deployed by the targeted uplink communication system, whose presence the BS intends to efficiently detect, e.g., for security purposes~\cite{RIS_safeguarding,WirelessCommIllegalRIS,metasurf_manip_onleg,RIS_Reciprocity,USE_IRS,RIS_Threat_OJCOM}.

\subsection{Channel Model}
A convenient wideband channel model for millimeter wave, and beyond, wireless communications is the delay-$d$ one presented in~\cite{delay_d_channel}, according to which, the \(N_{\rm bs}\times N_{\rm ue}\) MIMO channel between the BS and UE at each $d$-th sampling point, with $d=0,1,\ldots,D-1$, consists of \(L_{{\rm bs},{\rm ue}}\) propagation paths and is mathematically represented as follows: 
\color{black}
\begin{align}\label{eq:delay_d_channel_no_RIS}
    \mathbf{H}[d]\triangleq\sqrt{\frac{{{\rm PL}_{{\rm bs},{\rm ue}}}}{L_{{\rm bs},{\rm ue}}}}\sum_{\ell=1}^{L_{{\rm bs},{\rm ue}}}&\beta_{\ell}\mathbf{a}_{N_{\rm bs}}\left(\theta_{\ell}^{\rm A,bs},\phi_{\ell}^{\rm A,bs}\right)\mathbf{a}^{\herm}_{N_{\rm ue}}\left(\theta^{\rm D, ue}_{\ell}\right) \nonumber& \\
    &\times p\left(d T_s-\tau_{\ell}\right),&
\end{align}
where \(\theta^{s_1,s_2}_\ell\) and \(\phi^{s_1,s_2}_\ell\) denote the elevation and azimuth angles with respect to each $\ell$-th channel path ($\ell=1,2,\ldots,L_{{\rm bs},{\rm ue}}$), respectively, with the superscript $s_1$ taking the letter ``${\rm A}$'' for the case of angle of arrival and ``${\rm D}$'' for angle of departure. Superscript $s_2$ takes the string ``${\rm bs}$'' when implying the BS and the string ``${\rm ue}$'' for the UE case. We assume that all involved angles are uniformly distributed in \([0,2\pi]\). In addition, \(\mathbf{a}_{N_{\rm bs}}\left(\cdot,\cdot\right)\) and \(\mathbf{a}_{N_{\rm ue}}\left(\cdot\right)\) denote the \({N_{\rm bs}}\)-element BS and \({N_{\rm ue}}\)-element UE steering vectors, respectively, whereas \(\beta_{\ell}\sim \mathcal{CN}(0,1)\) is the complex gain of each $\ell$-th channel path, and ${{\rm PL}_{{\rm bs},{\rm ue}}}$ is the path loss. Finally, \(p(\cdot)\) is a pulse-shaping function for \(T_s\)-spaced signaling \cite[eq. (3)]{delay_d_channel}, and \(\tau_{\ell}\) represents the propagation delay for each \(\ell\)-th channel path, which is assumed to take values in the interval \([0, DT_s]\)~\cite{delay_d_channel}. 

To account for the impact of RISs in the wireless propagation environment, the linear cascaded channel model was first presented in~\cite{Huang_Reconfigurable_2019}, which has been shown via a first-principles coupled-dipole formalism~\cite{PhysFad} to reasonably approximate signal propagation in MIMO systems, especially with extensive number of antennas operating at high frequencies~\cite{rabault2024tacit}. To this end, considering \(R\) planar reflectors that are either fully or partially covered by RISs in the wireless environment, the end-to-end (${\rm e2e}$) MIMO channel between the BS and UE can be expressed as the superposition of the direct channel $\mathbf{H}[d]$ and the \(N_{\rm bs}\times N_{\rm ue}\) cascaded channel $\mathbf{H}_{\rm ref}[d]$. The latter models propagation from the UE towards the BS via the \(R\) planar reflectors, and can be expressed as follows:
\begin{align}\label{eq:cascade_channel}
        &\mathbf{H}_{\rm ref}[d] \triangleq  \sum_{r=1}^{R} \sqrt{\frac{{{\rm PL}_{{\rm bs},r}}{{\rm PL}_{r,{\rm ue}}}}{L_{{\rm bs},r}L_{r,{\rm ue}}}} \sum_{q=1}^{L_{{\rm bs},r}} \beta_{r,q}^{\rm bs}\mathbf{a}_{N_{\rm bs}}\left(\phi_{r,q}^{\rm A,bs},\theta_{r,q}^{\rm A,bs}\right)\nonumber\\
        & \times \mathbf{a}^{\herm}_{N_{\rm s},r}\left(\phi_{r,q}^{\rm D}, \theta_{r,q}^{\rm D}\right)\mathbf{S}_r[dT_s] \sum_{n=1}^{L_{r,{\rm ue}}}\beta_{r,n}^{\rm ue}\mathbf{a}_{N_{\rm s},r}\left(\phi_{r,n}^{\rm A},\theta_{r,n}^{\rm A}\right)\nonumber\\
        & \times \mathbf{a}_{N_{\rm ue}}^{\herm}\left(\theta_{r,n}^{{\rm D},{\rm ue}}\right)p\left(d T_s-\tau^{\rm bs}_{r,q}-\tau_{r,n}^{\rm ue}\right),
\end{align}
yielding the following expression for the ${\rm e2e}$ MIMO channel:
\begin{equation}\label{eq:e2e_delay_channel}
        \mathbf{H}_{\rm e2e}[d] \triangleq \mathbf{H}[d]+\mathbf{H}_{\rm ref}[d].
\end{equation}
In \eqref{eq:cascade_channel}, \(L_{{\rm bs},r}\) and \(L_{r,{\rm ue}}\) denote respectively the number of channel paths between each \(r\)-th planar reflector ($r=1,2,\ldots,R$) and the BS and between the UE and this reflector, whereas ${{\rm PL}_{{\rm bs},r}}$ and ${{\rm PL}_{r,{\rm ue}}}$ represent the respective channel path losses with the product \({{\rm PL}_{{\rm bs},r}}{{\rm PL}_{r,{\rm ue}}}\) implying the multiplicative path loss contributed in the signal propagation through each \(r\)-th reflector. Similar to the angles' definition in~\eqref{eq:delay_d_channel_no_RIS}, \(\theta_{r,x}^{s_1,s_2}\) and \(\phi_{r,x}^{s_1,s_2}\) denote respectively the elevation and azimuth angles with respect to each \(r\)-th surface reflector and each \(x\)-th channel path, where subscript \(x\) takes the values \(q=1,2,\ldots,L_{{\rm bs},r}\) for the channel between each \(r\)-th reflector and the BS, and the values \(n=1,2,\ldots L_{r,{\rm ue}}\) for the channel between the UE and each \(r\)-th reflector, while superscripts \(s_1\) and \(s_2\) take the same values as in \eqref{eq:delay_d_channel_no_RIS} with the only difference being that string \(s_2\) is empty when referring to any of the reflectors. Furthermore, \(\beta^{s_2}_{r,x}\sim \mathcal{CN}(0,1)\) and \(\tau^{s_2}_{r,x}\) represent the complex path gain and the propagation delay from each \(x\)-th channel path passing through each \(r\)-th reflector, whereas \(s_2\) takes the values  ``${\rm bs}$'' and ``${\rm ue}$'' when referring to the channel from the reflector to the BS or the UE to the reflector. Finally, similar to \(\tau_{\ell}\) in \eqref{eq:delay_d_channel_no_RIS}, the values \(\tau^{\rm bs}_{r,q} + \tau_{r,n}^{\rm ue}\,\forall\, r,q,n\) are assumed to lie in \([0, DT_s]\). 

\begin{remark}
    By assuming that: 1) the cascaded channel for each \(r\)-th reflector consists of a single path, i.e., \(L_{{\rm bs},r} = L_{r,{\rm ue}} = 1\,\forall r\); and 2) the surfaces are time invariant, i.e., \(\mathbf{S}_r[dTs]=\mathbf{S}_r[0]\,\forall d\), the expression for \(\mathbf{H}_{\rm ref}[d]\) given in~\eqref{eq:cascade_channel} can be incorporated into that for \(\mathbf{H}[d]\) in~\eqref{eq:delay_d_channel_no_RIS} by adding \(R\) more paths in the latter's summation.
\end{remark}
\begin{proof}
Starting from~\eqref{eq:cascade_channel} yields:
\begin{align}
   & \mathbf{H}_{\rm ref}[d] = \sum_{r=1}^{R} \sqrt{{{\rm PL}_{{\rm bs},r}}{{\rm PL}_{r,{\rm ue}}}}\beta_{r,1}^{\rm bs}\beta_{r,1}^{\rm ue} \mathbf{a}_{N_{\rm bs}}\!\left(\phi_{r,1}^{\rm A,bs},\theta_{r,1}^{\rm A,bs}\right) \nonumber\\
   & \hspace{1.39cm}\times \underbrace{\mathbf{a}^{\herm}_{N_{\rm s},r}\left(\phi_{r,1}^{\rm D}, \theta_{r,1}^{\rm D}\right)\mathbf{S}_r[dT_s]\mathbf{a}_{N_{\rm s},r}\!\left(\phi_{r,1}^{\rm A},\theta_{r,1}^{\rm A}\right)}_{\triangleq \alpha_r}\nonumber\\
   & \hspace{1.39cm}\times\mathbf{a}_{N_{\rm ue}}^{\herm}\!\left(\theta_{r,1}^{{\rm D},{\rm ue}}\right)p(d T_s-(\underbrace{\tau_{r,1}^{\rm bs}+\tau_{r,1}^{\rm ue}}_{\triangleq\tau_{r}^{\rm bs,ue}}))\Rightarrow \nonumber\\
   & \mathbf{H}_{\rm ref}[d] = \sum_{r=1}^{R} \sqrt{{{\rm PL}_{{\rm bs},r}}{{\rm PL}_{r,{\rm ue}}}}\beta_{r,1}^{\rm bs,r}\beta_{r,1}^{\rm r,ue} \alpha_r \mathbf{a}_{N_{\rm bs}}\!\left(\phi_{r,1}^{\rm A,bs},\theta_{r,1}^{\rm A,bs}\right)\nonumber\\
   &\hspace{1.39cm} \times \mathbf{a}_{N_{\rm ue}}^{\herm}\left(\theta_{r,1}^{{\rm D},{\rm ue}}\right)p\left(d T_s-\tau_{r}^{\rm bs,ue}\right).\label{eq:H_ref_to_H}
\end{align}
It can be seen that \(\mathbf{H}_{\rm ref}[d]\) in~\eqref{eq:H_ref_to_H} has the same form with \(\mathbf{H}[d]\) in~\eqref{eq:delay_d_channel_no_RIS}, with the only difference being some scalar multiplications. This difference can be eliminated by adding \(R\) more paths resulting from \(\mathbf{H}_{\rm ref}[d]\) in~\eqref{eq:H_ref_to_H} into the summation for \(\mathbf{H}[d]\) appearing in~\eqref{eq:delay_d_channel_no_RIS}.
\end{proof}

By modeling each \(r\)-th reflector as a surface with $N_{{\rm s},r}$ uniformly placed elements similar to a URA (and similar to the coupled-dipole formalism~\cite{PhysFad,rabault2024tacit}), with \(\mathbf{a}_{N_{\rm s},r}(\cdot,\cdot)\) denoting its steering vector, each matrix \(\mathbf{S}_r[dT_s]\in \mathbb{C}^{N_{{\rm s},r}\times N_{{\rm s},r}}\) in~\eqref{eq:e2e_delay_channel} includes the surface's reflection coefficients which can, in general, change in time; this particularly holds true for the parts of the surface covered by an RIS with time-varying phase/reflection configuration~\cite{RISsurvey2023}. We denote the \(r\)-th RIS's reflection vector as \(\boldsymbol{\phi}_r[d T_s]\in \mathbb{C}^{M_r \times 1}\) and the fixed reflection coefficient vector from the rest of the surface as \(\boldsymbol{\gamma}_r \in \mathbb{C}^{{(N_{{\rm s},r}}-M_r) \times 1}\), where \(N_{{\rm s},r}\geq M_r\). Hence, each \(\mathbf{S}_r[dTs]\) matrix appearing in \eqref{eq:cascade_channel} is modelled as a diagonal matrix containing the vectors \(\boldsymbol{\phi}[dT_s]\) and \(\boldsymbol{\gamma_r}\) in its diagonal \footnote{For the case of a beyond diagonal RIS \cite{BD_RIS}, \(\mathbf{S}_r[dT_s]\) can be extended to a non-diagonal matrix, by adding the entries corresponding to the coupling coefficients between RIS elements.}.


By taking the discrete Fourier transform of $\mathbf{H}_{\rm e2e}[d]$ in~\eqref{eq:e2e_delay_channel}, the ${\rm e2e}$ MIMO channel matrix between the BS and UE, and accordingly, the baseband received signal at each \(k\)-th subcarrier (\(k=0,1,\ldots,K-1\)) are given respectively as follows:
\begin{align}
    \boldsymbol{\mathcal{H}}_{\rm e2e}[k] &=\sum_{d=0}^{D-1}\mathbf{H}_{\rm e2e}[d]e^{-\jmath \frac{2 \pi k}{K}d}, \label{eq:channel_k_subcarrier} \\ 
    \mathbf{y}[k] &= \boldsymbol{\mathcal{H}}_{\rm e2e}[k]\mathbf{x}[k] + \mathbf{n}[k], \label{eq:received_signal}
\end{align}
where \(\mathbf{x}[k] \in \mathbb{C}^{N_{\rm ue}}\) represents the signal transmitted from the UE and \(\mathbf{n}[k]\sim \mathcal{CN}\left(\mathbf{0}_{N_{{\rm bs}}\times1}, \sigma^2 \mathbf{I}_{N_{\rm bs}}\right)\) models the additive white Gaussian noise, both referring to the \(k\)-th subcarrier.

\subsection{Problem Formulation}
The uplink MIMO communication between the UE and the BS takes place on a frame-by-frame basis through the transmission from the former of $M$ OFDM symbols $\mathbf{X}_m\triangleq[\mathbf{x}_m[1],\mathbf{x}_m[2],\ldots,\mathbf{x}_m[K]] \in \mathbb{C}^{N_{\rm ue}\times K}$ with $m=1,2,\ldots,M$ per frame We assume that each frame spans a coherent channel block (i.e., all channel paths remain the same throughout each frame transmission duration). Let the integer parameter $\mu\in[0,M]$ represent the time instant when the activity of one or more non-cooperative third-party RISs begins. It is assumed that both the value of $\mu$ and the characteristics of those RISs (e.g., size and number of elements, placement, phase
configuration codebook, reconfiguration frequency, and operation objective) are unknown to either the UE or the BS.

Our goal, in this paper, is to design a statistical test at the receiver side, i.e., at the BS, that processes the observation matrices  $\mathbf{Y}_1,\mathbf{Y}_2,\ldots,\mathbf{Y}_M$ per frame, with $\mathbf{Y}_m\triangleq[\mathbf{y}_m[1],\mathbf{y}_m[2],\ldots,\mathbf{y}_m[K]]\in\mathbb{C}^{N_{\rm bs} \times K}$ $\forall$$m$ via \eqref{eq:received_signal}, in an online manner to detect the presence of any non-cooperative RIS. More specifically, the BS designs a binary online decision rule that monitors all received OFDM symbols (data and/or pilots). In particular, for each $m$-th OFDM symbol per frame, it formulates the following binary variable: 
\begin{equation}
\label{eq:rule}
    s_{m,\mathbf{W}}=\begin{cases}
        1, \quad \text{highly probably RIS activity}\\
        0, \quad \text{otherwise}
    \end{cases},
\end{equation}
where $\mathbf{W}$ includes learnable parameters, e.g., a neural network. In the next section, we will elaborate on $s_{m,\mathbf{W}}$'s structure.

Let $\hat{\mu}_\mathbf{W}$ represent the first time where the latter BS rule in \eqref{eq:rule} outputs the value $1$ (i.e., RIS(s) detected). We define the expected RIS activity detection delay as follows: 
\begin{equation}
    \mathcal{P}_d(\hat{\mu}_{\mathbf{W}},\mu)\triangleq E\left[\hat{\mu}_\mathbf{W} -\mu\right], 
\end{equation}
as well as the false alarm rate as: 
\begin{equation}
    \mathcal{P}_f(\hat{\mu}_\mathbf{W},\mu)\triangleq{\rm Pr}\left[\hat{\mu}_\mathbf{W} < \mu | \mu \rightarrow \infty \right].
\end{equation}
Note that the detection delay is not measured in units of time (e.g., seconds), but in the number of OFDM symbols received under RIS(s)' presence without any alert. Our RIS activity detection objective is expressed via the following optimization problem with respect to the $\mathbf{W}$ parameters:
\begin{equation*}
\mathcal{OP}_1: \underset{{\rm \hat{\mu}_\mathbf{W} }}{\min} \mathcal{P}_d(\hat{\mu}_\mathbf{W},\mu) \quad \text{s.t.} \quad \mathcal{P}_f(\hat{\mu}_\mathbf{W},\mu) \leq F,
\end{equation*}
where $F$ is a threshold depending on the MIMO communication metric. It is noted that our detection objective can tolerate higher false alarm rates, even if this leads to shorter delays.

\section{Proposed dSVDD-Based Scan $B$-statistic}
With the advances in graphics and tensor processing units, pre-trained large neural networks can rapidly process high dimensional input data. To this end, enhancing existing statistical tests with neural networks emerges as a promising solution. In addition, there has recently been a lot of progress in pruning methods (see, e.g.,~\cite{stateNetPrun} and references therein), which can further speed up inference time, in our case, the detection of activity from one or more non-cooperative RISs. 

In this section, we present a novel unsupervised distribution-free change point detection method, which is termed as dSVDD-scan $B$-test, for solving $\mathcal{OP}_1$. Our RIS activity detection learning method does make any assumption on the structure of the post-change data, and its neural network does not require training with observations affected by non-cooperative RISs. In addition, while our method deploys a variant of the dSVDD model in~\cite{dsvdd} that is typically used for anomaly detection, we do not attempt to tackle a simple static anomaly detection problem. The stochasticity in the transmitted OFDM symbols and the noise indicates that there will probably be abnormal signal observations at the BS even when UE transmissions are not influenced by the activity of non-cooperative RIS(s).

\subsection{The dSVDD Artificial Neural Network Architecture}
We use the dSVDD model~\cite{dsvdd} to extract representative and low dimensional features from the BS observations, and then, feed those features to the scan $B$-statistic~\cite{scanB}. As it will be experimentally verified in the next section, this can greatly enhance the RIS activity detection performance. More specifically, before passing each $m$-th received OFDM symbol $\mathbf{Y}_m$ to the neural network, we flatten it and then stack its real and imaginary parts into a $2N_{\rm bs}K \times 1$ real-valued vector $\mathbf{v}_m$.

The dSVDD model, which is a deep learning variant of the SVDD model \cite{svdd}, works by first transforming each input $\mathbf{v}_m$ to a latent lower dimensional representation $\boldsymbol{\lambda}_m\triangleq g_\mathbf{W} (\mathbf{v}_m) \in \mathbb{R}^{N_{\rm LAT}\times1}$, where $\mathbf{W}$ denotes the parameters of the neural network and $N_{\rm LAT}$ indicates the dimension of the latent space. The model's outputs $\boldsymbol{\lambda}_m$'s for all normal inputs $\mathbf{v}_m$'s (i.e., observations unaffected by any RIS activity) will be bounded inside an $N_{\rm LAT}$-dimensional sphere with center $\mathbf{c} \in \mathbb{R}^{N_{\rm LAT}\times1}$ and radius $R$. We define the test for RIS activity within the wireless environment of the MIMO communication system as: 
\begin{equation}\label{eq:sigma_test}
   \sigma_\mathbf{W} (\mathbf{v}_m) \triangleq ||\boldsymbol{\lambda}_m - \mathbf{c}||_2^2 \geq R,
\end{equation} 
where the learnable function $\sigma_\mathbf{W}(\cdot)$ is known as the anomaly score. In the event that \eqref{eq:sigma_test} holds true (i.e., $\boldsymbol{\lambda}_m$ lies outside the $R$-radius sphere), the $m$-th received OFDM symbol is classified as an anomaly (i.e., affected by unwanted RIS activity).

To train our modified dSVDD model hosted at the BS, we construct a large dataset $\mathcal{E}$ of actually received OFDM sysmbols (i.e., similar to \eqref{eq:received_signal} for large $M$). This set should include signals unaffected by any unintended RIS activity, constituting in this way the normal inputs to the model. Otherwise, the model will be trained with observations including unintended RIS activity, which will prevent it from detecting similar activity during its deployment phase. Our dSVDD model is then trained over $\mathcal{E}$ to solve the following optimization problem~\cite{dsvdd}:
\begin{equation*}
\mathcal{OP}_2: \underset{R,\mathbf{W}}{\min}\,R^2 +\frac{1}{\lambda_1 |\mathcal{E}|} \sum_{\mathbf{v}_m \in \mathcal{E}}\!\! \max\{0,\sigma_\mathbf{W}(\mathbf{v}_m)-R^2\}+\lambda_2 \text{r}(\mathbf{W}),
\end{equation*}
where $\text{r}(\cdot)$ is a regularization function on the model's parameters (e.g., a weight decay regularizer), and $\lambda_1$ and $\lambda_2$ are hyperparameters. The former hyperparameter controls the number of training data points allowed to be outside the sphere (i.e., the number of RIS-affected observations (anomalies) being present in $\mathcal{E}$), while the latter deals with the strength of the regularizer in $\mathcal{OP}_2$. The center $\mathbf{c}$ in~\eqref{eq:sigma_test}, and consequently in the problem formulation,  is computed by averaging all latent representations within $\mathcal{E}$, whereas the radius $R$ will be learned together with the model's parameters $\mathbf{W}$. To jointly optimize these variables, we treat them as a single variable and deploy a stochastic gradient descent variant, like the Adam optimizer~\cite{adam}, paired with backpropagation. We will denote by  $\mathbf{W}^{*}$ the parameters of the trained model.

\subsection{RIS Activity Test Statistic}
The scan $B$-statistic is typically paired with a kernel function which can make its computation significantly slower~\cite{scanB}. We will instead use the trained model $\mathbf{W}^*$ to extract low dimensional information, and then, perform the RIS activity test on this information. The key motivation here lies in the fact that the anomaly score's distribution will change once activity from any non-cooperative RIS takes place.

Suppose that the integer value $m_0$ indicates that, up to the $m_0$-th received OFDM symbol, the uplink MIMO communication takes place in the abscence of any non-cooperative RIS activity. To this end, the resulting reference observations $\mathbf{Y}_1,\mathbf{Y}_2,\ldots,\mathbf{Y}_{m_0}$ are first transformed to $\sigma_\mathbf{W^*}(\mathbf{v}_1),\sigma_\mathbf{W^*}(\mathbf{v}_2),\ldots,\sigma_\mathbf{W^*}(\mathbf{v}_{m_0})$ via the trained neural network.
The reference sequence is then split into the $m_w$ blocks $\mathbf{s}_1,\mathbf{s}_2,\ldots,\mathbf{s}_{m_w}$ of length $W$ each, where $m_w\triangleq\frac{m_0}{W}$. For instance, $\mathbf{s}_1=[\sigma_\mathbf{W^*}(\mathbf{v}_1),\sigma_\mathbf{W^*}(\mathbf{v}_2),\ldots,\sigma_\mathbf{W^*}(\mathbf{v}_{W})]$.
Then, for each $m$-th received OFDM symbol $\mathbf{Y}_m$  with $m>m_0$, $\sigma_\mathbf{W^*}(\mathbf{v}_m)$ is obtained that is followed by the generation of a vector including the most recent $W$ observations, i.e.:
\begin{equation*}
    \mathbf{s}_{m-W:m}=[\sigma_\mathbf{W^*}(\mathbf{v}_{m-W}),\sigma_\mathbf{W^*}(\mathbf{v}_{m-W+1}),\ldots,\sigma_\mathbf{W^*}(\mathbf{v}_m)].
\end{equation*}
For each of these blocks, we wish to examine whether it differs significantly from each of the latter reference blocks. 
Therefore, we define the scan $B$-statistic for each $m$-th observation as:
\begin{equation}
 z_m\triangleq m_w^{-1}\sum_{n=1}^{m_w}{\rm MMD}(\mathbf{s}_n,\mathbf{s}_{m-W:m}),
\end{equation}
where ${\rm MMD}(\cdot)$ is an unbiased estimator of the maximum mean discrepancy~\cite{scanB}. 

Let $k(\cdot,\cdot)$ be a kernel function, e.g., the linear kernel or the Gaussian radial basis function kernel, based on which we define the following function:
\begin{align}\label{eq:h_function}
h(x_i,x_j,y_i,y_j)\triangleq& k(x_i,x_j)+k(y_i,y_j) \nonumber & \\&-k(x_i,y_j)-k(x_j,y_i).
\end{align}
The maximum mean discrepancy estimator between any post change block $\mathbf{s}_{m-W:m}$ and the $n$-th reference block $\mathbf{s}_n$ can be obtained by averaging the function in \eqref{eq:h_function} between all pairs in the former block and all pairs in the latter block, i.e., \cite{GrettonTwosample}:
\begin{align}
&{\rm MMD}(\mathbf{s}_n,\mathbf{s}_{m-W:m})= \frac{1}{W(W-1)}\nonumber& 
\\
&\times\sum_{i \neq j}^{W}h(\mathbf{s}_n[i],\mathbf{s}_n[j],\mathbf{s}_{m-W:m}[i],\mathbf{s}_{m-W:m}[j]),&
\end{align}
where $\mathbf{s}_n[i]$ denotes the $i$-th ($i=1,2,\ldots,W$) element of $\mathbf{s}_n$; for instance, $\mathbf{s}_1[2]=\sigma(\mathbf{v}_2)$. Recall that we have assumed that $\mathbf{W}^*$ extracts all the necessary information from the model's input vectors, thus, a linear kernel function suffices. To this end, the decision rule of \eqref{eq:rule} for the RIS activity is deduced to: 
\begin{equation}\label{eq:final_rule}
 s_{m,\mathbf{W}^*}=\begin{cases}
     1, \quad \text{if  } z_m \geq \theta\\
     0, \quad \text{otherwise}
 \end{cases},
\end{equation}
where $\theta$ is a communication-dependent threshold.

\section{Numerical Results and Discussion}
\subsection{Simulation Setup}
The intended uplink MIMO OFDM communications take place in a wireless environment with $R=2$ planar reflectors.
We have set $N_{\rm ue}=10$, $N_{\rm bs}=20$, the cyclic prefix length $D=128$, the number of propagation paths for all links as $L_{\rm bs}=L_{\rm ue}=L_{{\rm bs},r}=L_{r,{\rm ue}}=4$ for $r=1$ and $2$, as well as $N_{{\rm s},1}=N_{{\rm s},2}=400$. 
The transmitted OFDM symbols were independent and identically distributed according to the standard complex Gaussian distribution. We have used the pulse shaping function described in \cite[Chapter 2]{heath2018foundations} for the channel model, and varied the number of subcarriers $K$ from $64$ to $512$. 
\begin{figure*}
    \centering
    \begin{subfigure}{.3\textwidth}
        \centering
\includegraphics[width=\textwidth]{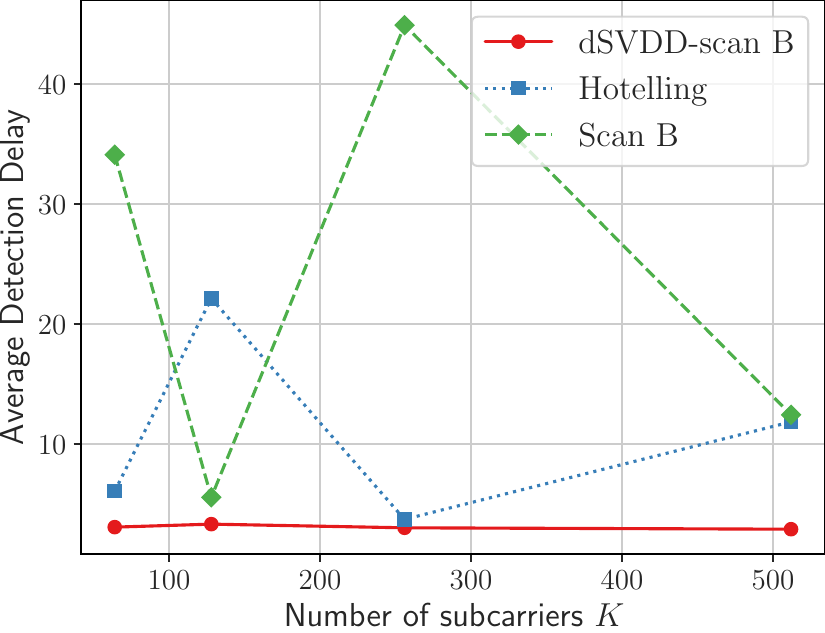}
        \caption{Average Detection Delay.}\label{subfig:delay}
    \end{subfigure} \hspace{0.25cm}
    \begin{subfigure}{.3\textwidth}
        \centering        \includegraphics[width=\textwidth]{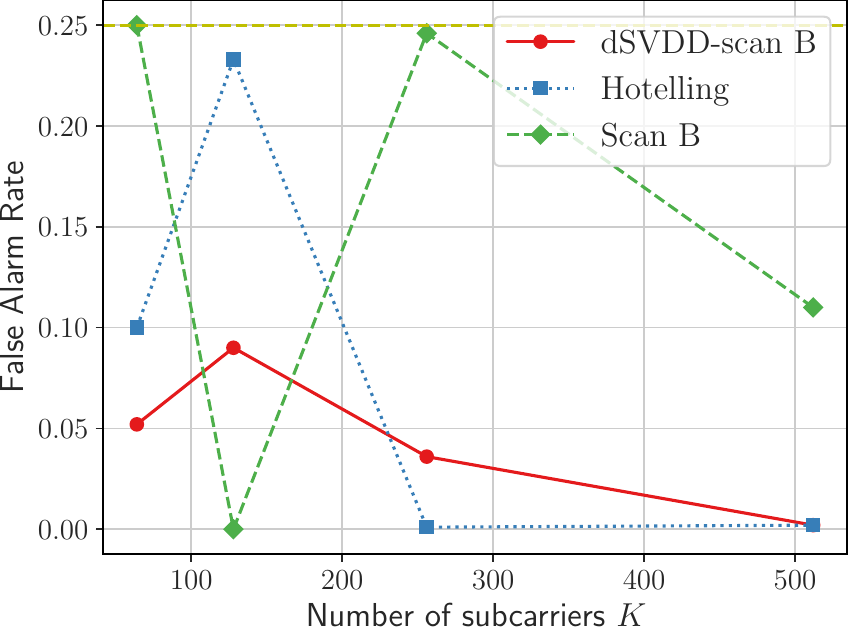}
        \caption{Empirical false alarm rate.}\label{subfig:far}
    \end{subfigure} \hspace{0.25cm}
    \begin{subfigure}{.3\textwidth}
        \centering
\includegraphics[width=\textwidth]{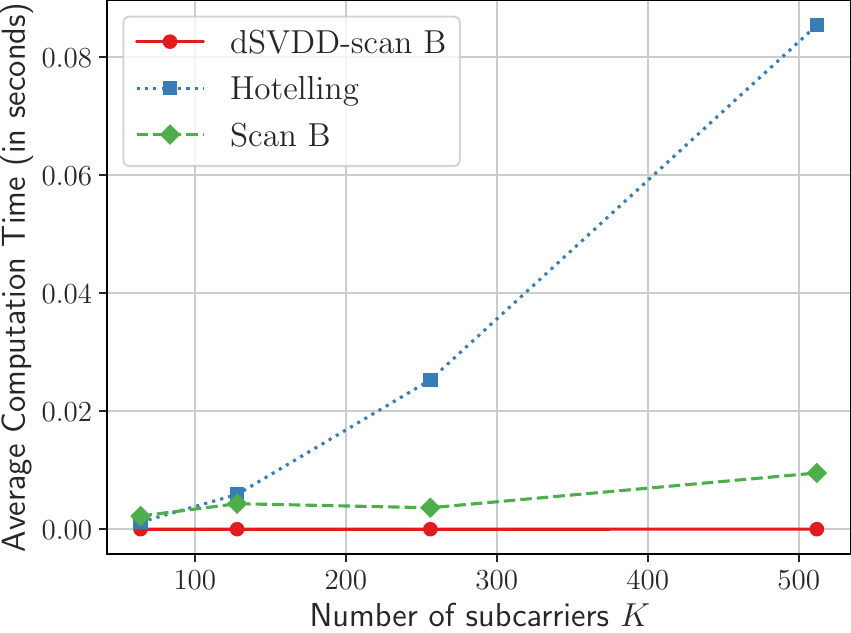}
        \caption{Average computation time.}\label{subfig:time}
    \end{subfigure}
    \caption{RIS detection performance for the first communication scenario at $8$ GHz with $R=2$ planar reflectors.}\label{fig:res1} 
\end{figure*}

We have considered the following two communication scenarios: \textit{i}) Central frequency at $3.5$ GHz, and \textit{ii}) Central frequency at $28$ GHz, with   both the two planar reflectors placed in between the BS and UE at the distance of $3$ meters. 
Both surfaces were simulated as fully static reflectors (i.e., $M_1=M_2=0$), with their reflection coefficient given as $\boldsymbol{\gamma}_2=\boldsymbol{\gamma}_1 = \mathbf{1}^{N_{\rm s,1}\times1} \delta$, where $\delta$ is uniformly distributed in $[0.68,0.71]$; the latter values for $\delta$ at $28$ GHz correspond to indoor dry wall environments~\cite{ReflectionCoeff}. For the environment with the $3.5$ GHz communication setup, we have set $\delta=1$. In addition, we have set $m_0=100$ and considered that, after the OFDM symbol with the index $\mu=150$ has arrived at the BS, the second reflector is coated with an RIS covering its entire surface (i.e., $M_2=400$). The first reflector was assumed to be free of any RIS throughout the whole duration of interest. A maximum horizon of $M=500$ OFDM symbols was considered for the test statistic to detect the RIS. If the BS could not detect the RIS within this horizon window, we assumed that the designed statistic has missed it. In order to be consistent with most of the available RIS prototypes~\cite{RIS_challenges}, 
we assumed that the RIS is equipped with $1$-bit phase resolution metamaterials. In particular, during the transmission of each $m$-th OFDM symbol with $m\geq\mu$, each RIS element's value was selected uniformly from $\{0,\pi\}$.

\subsection{Implementation Details and Benchmark Schemes}
The training dataset $\mathcal{E}$ consisted of $60000$ observations via \eqref{eq:received_signal}, and the reference sequence length was set as $m_0=100$. For benchmark, we have implemented the classical Hotelling test and the scan $B$-statistic~\cite{scanB} with a linear kernel performed on the flattened observation arrays $\mathbf{v}_m$ $\forall$$m$. The window length for all test statistics was set as $W=5$.

We have built our modified dSVDD architecture using the pytorch framework \cite{pytorch}. Our artificial neural network consisted of $4$ linear layers, each with a different number of units, which were empirically selected. The number of output units of each layer was $64\times20$, $64\times10$, $64\times5$, and $32$. Each layer was followed by a leaky rectified linear unit activation function. Our network's training was performed using the Adam~\cite{adam} optimizer with a weight decay of $10^{-4}$ for regularization. Our simulations were conducted on a desktop computer with an $11$-th generation Intel
Core i7 CPU, $32$ GB RAM, and an Nvidia RTX 3080 ($10$ GB
VRAM) GPU.

\subsection{Detection Performance Evaluation}
We have evaluated the average detection delay, the empirical false alarm rate, and the average computation time for all studied RIS activity detection schemes using $1000$ Monte-Carlo episodes each with $M$ OFDM symbol transmissions. The latter metric quantifies the average time needed to compute the test statistic during per received OFDM symbol. The communication-dependent threshold $\theta$ in \eqref{eq:final_rule} for our dSVDD-scan $B$ test and the benchmark test statistics was selected such that the false alarm rate was at most $F=0.25$.

\begin{figure}
    \centering
    \includegraphics[width=0.3\textwidth]{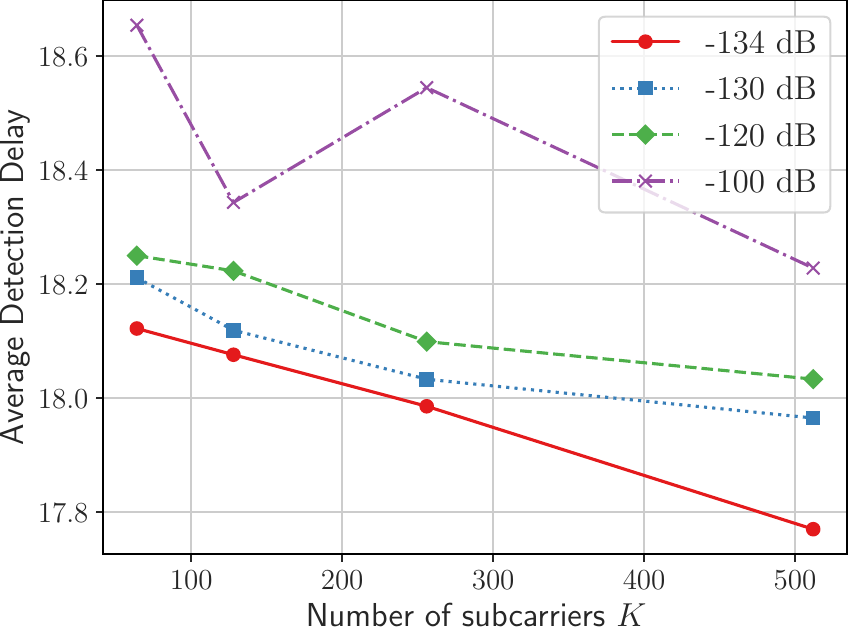}
    \caption{RIS detection delay performance for the second communication scenario at $28$ GHz with $R=2$ planar reflectors considering different noise levels.}
    \label{fig:resNormal}
\end{figure}

Figure~\ref{fig:res1} includes all detection results with all three compared test statistics for the first
communication scenario at $3.5$ GHz, while Fig.~\ref{fig:resNormal} illustrates the average RIS detection delay performance for the second
communication scenario at $28$ GHz for different noise levels, but only for the proposed dSVDD-scan $B$ test since RIS detection with the two benchmarks was impossible. As depicted in Fig.~\ref{subfig:delay} for the noise level $-120$~dB, the proposed dSVDD-scan $B$-statistic achieves the shortest delay for all $K$ values. In fact, we have seen that, for our method, the delay tends to decrease for large $K$. In contrast, it is shown that both the Hotelling test and the scan $B$-test have rather unstable performance. For some values of $K$, they achieve pretty small delays and, for some others, very large. A similar trend for all three test statistics, and their comparative behavior, can be seen in Fig.~\ref{subfig:far} for the false alarm rate. Clearly, the proposed test statistic always achieves false alarm rates significantly smaller than $F$. Finally, in Fig.~\ref{subfig:time}, it is demonstrated that our dSVDD-scan $B$ test requires the smaller computation time, with its average run time being almost identical for all considered values of $K$, unlike the conventional scan $B$-statistic and the Hotelling test. In fact, it can be observed that the latter approach's run time increases almost exponentially with increasing $K$ values.

For the demanding communication scenario considered in Fig.~\ref{fig:resNormal}, we have found that the proposed dSVDD-scan $B$-statistic is always less than the threshold of $F=0.25$. In addition, it is shown in the figure that our approach is capable to detect the presence of an operating RIS in less than $19$ OFDM symbol transmissions, for all simulated noise levels. As depicted, and as expected, the RIS detection delay decreases with increasing $K$ values.

\section{Conclusion}
In this paper, we presented a novel distribution-free change point detection statistic for high dimensional data that does not require a pre-change reference block. Our dSVDD-scan $B$ test was applied to detect non-cooperative activity from one or more RISs with unknown characteristics impacting a point-to-point MIMO OFDM communication system. We also presented a novel wideband channel model incorporating RISs as well as non-reconfigurable stationary surfaces. Our performance evaluations demonstrated that the proposed detection scheme can quickly detect unwanted RIS activity, while being much more computationally efficient than existing detection methods.

\bibliographystyle{IEEEtran}
\bibliography{references}

\begin{thebibliography}{10}
\providecommand{\url}[1]{#1}
\csname url@samestyle\endcsname
\providecommand{\newblock}{\relax}
\providecommand{\bibinfo}[2]{#2}
\providecommand{\BIBentrySTDinterwordspacing}{\spaceskip=0pt\relax}
\providecommand{\BIBentryALTinterwordstretchfactor}{4}
\providecommand{\BIBentryALTinterwordspacing}{\spaceskip=\fontdimen2\font plus
\BIBentryALTinterwordstretchfactor\fontdimen3\font minus \fontdimen4\font\relax}
\providecommand{\BIBforeignlanguage}[2]{{%
\expandafter\ifx\csname l@#1\endcsname\relax
\typeout{** WARNING: IEEEtran.bst: No hyphenation pattern has been}%
\typeout{** loaded for the language `#1'. Using the pattern for}%
\typeout{** the default language instead.}%
\else
\language=\csname l@#1\endcsname
\fi
#2}}
\providecommand{\BIBdecl}{\relax}
\BIBdecl

\bibitem{RISsurvey2023}
E.~Basar \emph{et~al.}, ``Reconfigurable intelligent surfaces for {6G}: {E}merging applications and open challenges,'' \emph{IEEE Veh. Technol. Mag.}, vol.~19, no.~3, pp. 27--47, 2024.

\bibitem{6Gdisac}
E.~Calvanese~Strinati \emph{et~al.}, ``Towards distributed and intelligent integrated sensing and communications for {6G} networks,'' \emph{arXiv preprint:2402.11630}, 2024.

\bibitem{AlexandroStylianoPervasive}
G.~C. Alexandropoulos \emph{et~al.}, ``Pervasive machine learning for smart radio environments enabled by reconfigurable intelligent surfaces,'' \emph{Proc. IEEE}, vol. 110, no.~9, pp. 1494--1525, 2022.

\bibitem{over_the_air_RIS}
W.~Ni \emph{et~al.}, ``Integrating over-the-air federated learning and non-orthogonal multiple access: What role can {RIS} play?'' \emph{IEEE Trans. Wirel. Commun.}, vol.~21, no.~12, pp. 10\,083--10\,099, 2022.

\bibitem{RIS_loc}
K.~Keykhosravi \emph{et~al.}, ``Leveraging {RIS}-enabled smart signal propagation for solving infeasible localization problems,'' \emph{IEEE Veh. Technol. Mag.}, vol.~18, no.~2, pp. 20--28, 2023.

\bibitem{RIS_isac}
S.~P. Chepuri \emph{et~al.}, ``Integrated sensing and communications with reconfigurable intelligent surfaces,'' \emph{IEEE Signal Process. Mag.}, vol.~40, no.~6, pp. 41--62, 2023.

\bibitem{RIS_safeguarding}
G.~C. Alexandropoulos \emph{et~al.}, ``Safeguarding {MIMO} communications with reconfigurable metasurfaces and artificial noise,'' in \emph{Proc. {IEEE ICC}}, Montreal, Canada, 2021.

\bibitem{RIS_Threat_OJCOM}
------, ``Counteracting eavesdropper attacks through reconfigurable intelligent surfaces: A new threat model and secrecy rate optimization,'' \emph{IEEE Open J. Commun. Society}, vol.~4, no.~3, p. 1285–1302, 2023.

\bibitem{WirelessCommIllegalRIS}
Y.~Wang \emph{et~al.}, ``Wireless communication in the presence of illegal reconfigurable intelligent surface: Signal leakage and interference attack,'' \emph{IEEE Wireless Commun.}, vol.~29, no.~3, pp. 131--138, 2022.

\bibitem{metasurf_manip_onleg}
H.~Alakoca \emph{et~al.}, ``Metasurface manipulation attacks: Potential security threats of {RIS}-aided {6G} communications,'' \emph{IEEE Commun. Mag.}, vol.~61, no.~1, pp. 24--30, 2023.

\bibitem{RIS_Reciprocity}
H.~Wang and A.~L. Swindlehurst, ``Channel reciprocity attacks using intelligent surfaces with non-diagonal phase shifts,'' \emph{arXiv preprint arXiv:2309.11665}, 2022.

\bibitem{USE_IRS}
S.~Sarp \emph{et~al.}, ``Use of intelligent reflecting surfaces for and against wireless communication security,'' in \emph{Proc. IEEE 5G World Forum}, 2021.

\bibitem{RIS_challenges}
G.~C. Alexandropoulos \emph{et~al.}, ``{RIS}-enabled smart wireless environments: Deployment scenarios, network architecture, bandwidth and area of influence,'' \emph{EURASIP J. Wireless Commun. and Netw.}, vol. 2023, no.~1, pp. 1--38, 2023.

\bibitem{RIS_THz_terrameta}
S.~Matos \emph{et~al.}, ``Reconfigurable intelligent surfaces for {THz}: {H}ardware impairments and switching technologies,'' in \emph{Proc. ICEAA}, Lisbon, Portugal, 2024.

\bibitem{delay_d_channel}
A.~Alkhateeb and R.~W. Heath, ``Frequency selective hybrid precoding for limited feedback millimeter wave systems,'' \emph{IEEE Trans. Commun.}, vol.~64, no.~5, pp. 1801--1818, 2016.

\bibitem{PhysFad}
R.~Faqiri \emph{et~al.}, ``{PhysFad}: Physics-based end-to-end channel modeling of {RIS}-parametrized environments with adjustable fading,'' \emph{IEEE Trans. Wireless Commun.}, vol.~22, no.~1, pp. 580--595, 2023.

\bibitem{changePointSurveys}
L.~Xie \emph{et~al.}, ``Sequential (quickest) change detection: Classical results and new directions,'' \emph{IEEE J. Sel. Areas Inf. Theory}, vol.~2, no.~2, pp. 494--514, 2021.

\bibitem{svdd}
D.~Tax and R.~Duin, ``Support vector data description,'' \emph{Mach. Learn.}, vol.~54, p. 45–66, 2004.

\bibitem{dsvdd}
L.~Ruff \emph{et~al.}, ``Deep one-class classification,'' in \emph{Proc. ICML}, Jul. 2018.

\bibitem{scanB}
S.~Li \emph{et~al.}, ``Scan {$B$}-statistic for kernel change-point detection,'' \emph{Sequential Analysis}, vol.~38, no.~4, pp. 503--544, 2019.

\bibitem{Huang_Reconfigurable_2019}
C.~Huang \emph{et~al.}, ``Reconfigurable intelligent surfaces for energy efficiency in wireless communication,'' \emph{IEEE Trans. Wireless Commun.}, vol.~18, no.~8, pp. 4157--4170, Aug. 2019.

\bibitem{rabault2024tacit}
A.~Rabault \emph{et~al.}, ``On the tacit linearity assumption in common cascaded models of {RIS}-parametrized wireless channels,'' \emph{IEEE Trans. Wireless Commun.}, early access, 2024.

\bibitem{BD_RIS}
D.~Wijekoon \emph{et~al.}, ``Physically-consistent modeling and optimization of non-local {RIS}-assisted multi-user {MIMO} communication systems,'' \emph{ar{X}iv preprint: 2406.05617}, 2024.

\bibitem{stateNetPrun}
D.~Blalock \emph{et~al.}, ``What is the state of neural network pruning?'' \emph{Proc. MLSys}, 2020.

\bibitem{adam}
D.~P. Kingma and J.~Ba, ``Adam: A method for stochastic optimization,'' \emph{ar{X}iv preprint:1412.6980s}, 2017.

\bibitem{GrettonTwosample}
A.~Gretton \emph{et~al.}, ``A kernel two-sample test,'' \emph{J. Mach. Learn. Res.}, vol.~13, no.~25, pp. 723--773, 2012.

\bibitem{heath2018foundations}
R.~W. Heath and A.~Lozano, \emph{Foundations of MIMO Communication}.\hskip 1em plus 0.5em minus 0.4em\relax Cambridge University Press, 2018.

\bibitem{ReflectionCoeff}
H.~Zhao \emph{et~al.}, ``$28$ {G}hz millimeter wave cellular communication measurements for reflection and penetration loss in and around buildings in {N}ew {Y}ork city,'' in \emph{Proc. {IEEE} {ICC}}, Budapest, Hungary, 2013.

\bibitem{pytorch}
A.~Paszke \emph{et~al.}, ``Pytorch: An imperative style, high-performance deep learning library,'' in \emph{Proc. NeurIPS}, 2019.

\end{thebibliography}
\vfill

\end{document}